\documentclass[aps,prb,twocolumn,superscriptaddress]{revtex4}

\usepackage{graphicx}
\usepackage[german,english]{babel}
\usepackage{amssymb}
\usepackage{amsmath}
\usepackage{subfigure}

\begin{document}
\title{Double Counting in LDA+DMFT -- The Example of NiO}

\author{M. Karolak}
\affiliation{I. Institut f{\"u}r Theoretische Physik, Universit{\"a}t Hamburg, Jungiusstra{\ss}e 9, D-20355 Hamburg, Germany}
\email{mkarolak@physnet.uni-hamburg.de}
\author{G. Ulm}
\affiliation{I. Institut f{\"u}r Theoretische Physik, Universit{\"a}t Hamburg, Jungiusstra{\ss}e 9, D-20355 Hamburg, Germany}
\author{T. O. Wehling}
\affiliation{I. Institut f{\"u}r Theoretische Physik, Universit{\"a}t Hamburg, Jungiusstra{\ss}e 9, D-20355 Hamburg, Germany}
\author{V. Mazurenko}
\affiliation{Theoretical Physics and Applied Mathematic Department, Urals State Technical University, 620002, Mira street 19, Yekaterinburg, Russia}
\author{A. Poteryaev}
\affiliation{Institute of Metal Physics, Russian Academy of Sciences, 620041 Yekaterinburg GSP-170, Russia}
\author{A. I. Lichtenstein}
\affiliation{I. Institut f{\"u}r Theoretische Physik, Universit{\"a}t Hamburg, Jungiusstra{\ss}e 9, D-20355 Hamburg, Germany}



\date{\today}

\begin{abstract}

An intrinsic issue of the LDA+DMFT approach is the so called double counting of interaction terms. How to choose the double-counting potential in a manner that is both physically sound and consistent is
unknown. We have conducted an extensive study of the charge transfer system NiO in the LDA+DMFT framework using quantum Monte
Carlo and exact diagonalization as impurity solvers. By explicitly
treating the double-counting correction as an adjustable parameter
we systematically investigated the effects of different choices for the
double counting on the spectral function. Different methods for fixing
the double counting can drive the result from Mott insulating to almost
metallic. We propose a reasonable scheme for the determination of double-counting corrections for insulating systems.                                     
                                                                                                                                                                                                                                      
\end{abstract}

\maketitle

\section{Introduction}
\label{introduction}

The combination of the density functional theory (DFT/LDA), a model Hamiltonian and the dynamical
mean field approximation (DMFT) \cite{dmft_review96}, a methodology commonly referred to as
LDA+DMFT, is to date one of the best approaches for the realistic description of strongly
correlated electron systems \cite{anisimov_dmft,lda++}.
While density functional theory does not include all the interactions between strongly correlated $d$ or $f$ electrons, it captures some portion of them through the Hartree and exchange-correlation terms. By introduction of a model Hamiltonian into the calculations one tries to account for as much of the interactions as possible through the Coulomb interaction matrix of the impurity model. This ultimately leads to the problem that some contributions to the interaction are included twice. This has to be explicitly compensated by adding a shift in the chemical potential of the correlated orbitals to the Hamiltonian, leading to the prominent issue of double counting. 
The LDA+DMFT Hamiltonian can be written as follows
\begin{eqnarray*}
H=&&H_{LDA}-H_{dc}+\\
\\
&&+\tfrac{1}{2}\hspace*{-15pt}\sum_{i, \sigma\sigma^\prime, m m^\prime m^{\prime\prime} m^{\prime\prime\prime}}
\hspace*{-15pt}U_{m m^\prime m^{\prime\prime}m^{\prime\prime\prime}}c^{\dagger}_{im\sigma}
c^{\dagger}_{im^\prime\sigma^\prime}c_{im^{\prime\prime\prime}\sigma}c_{im^{\prime\prime}\sigma}
\end{eqnarray*} 
where $H_{LDA}$ is the LDA Hamiltonian, $c^{\dagger}_{im\sigma}$ creates a particle with spin $\sigma$ in a localized orbital $m$ at site $i$  and $U_{m m^\prime m^{\prime\prime}m^{\prime\prime\prime}}$ is the Coulomb interaction matrix between localized orbitals. Above Hamiltonian contains the double-counting correction
\[H_{dc}=\mu_{dc}\sum_{m,\sigma} n_{m,\sigma},\]
where $n_{m,\sigma}=c^{\dagger}_{m\sigma}c_{m\sigma}$ and $\mu_{dc}$ is the double-counting potential.
How to choose the double-counting potential in a manner that is physically sound and consistent is unknown and systematic investigations of the effects of the double counting in LDA+DMFT on the spectrum are seldom performed. In the work presented here we attempt to shed some light on the double-counting problem using the example of nickel oxide (NiO). In recent years a number of authors applied the LDA+DMFT method in different flavors to this system generating a body of promising results \cite{ren_nio,kunes_nio1,kunes_nio2,korotin_wannier}.
\section{NiO -- a charge transfer system}

\begin{figure}[t]
\centering
 \begin{minipage}{0.49\linewidth}
\centering
 \includegraphics[width=\linewidth]{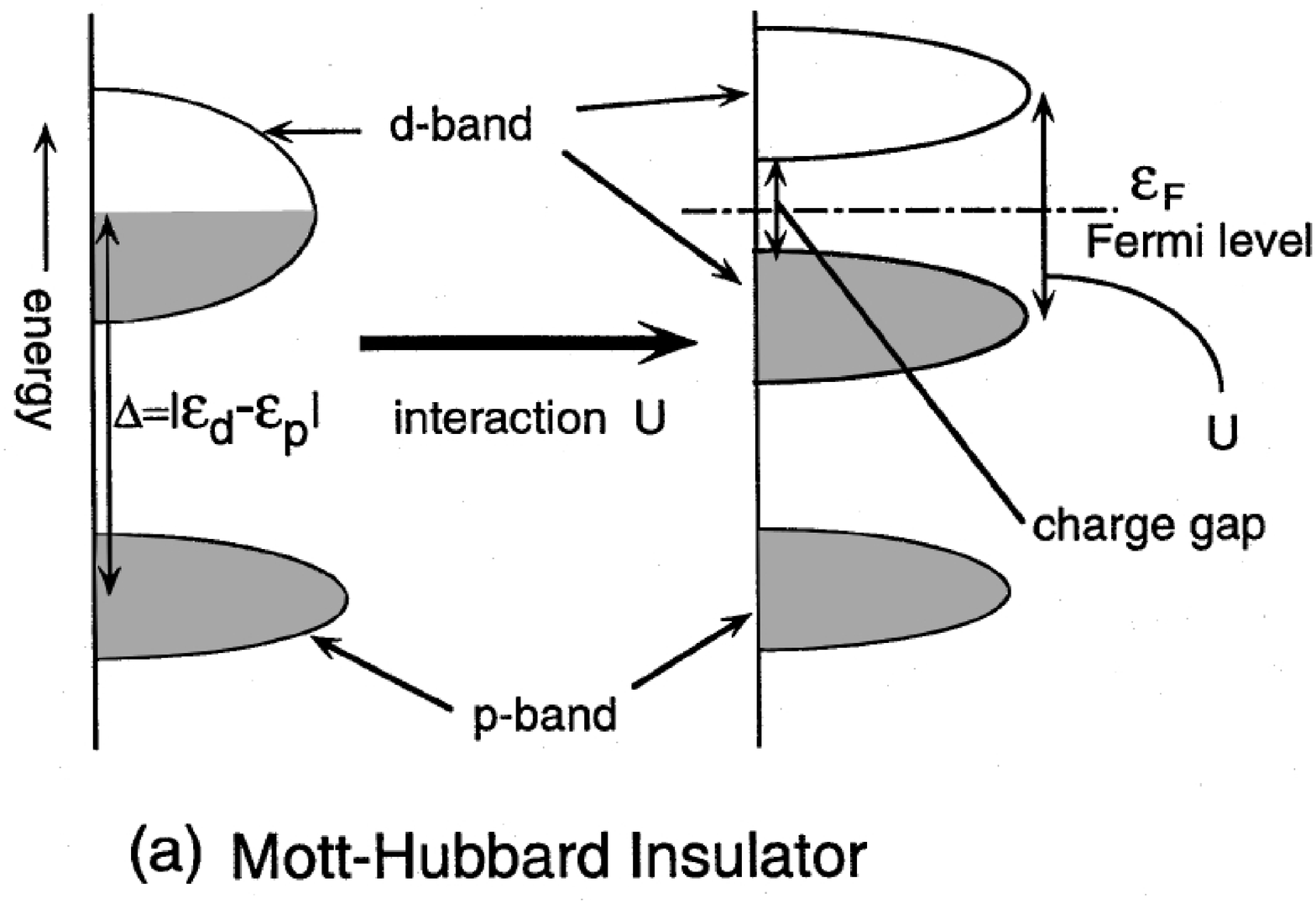}
 \end{minipage}
 \begin{minipage}{0.49\linewidth}
  \centering
   \includegraphics[width=\linewidth]{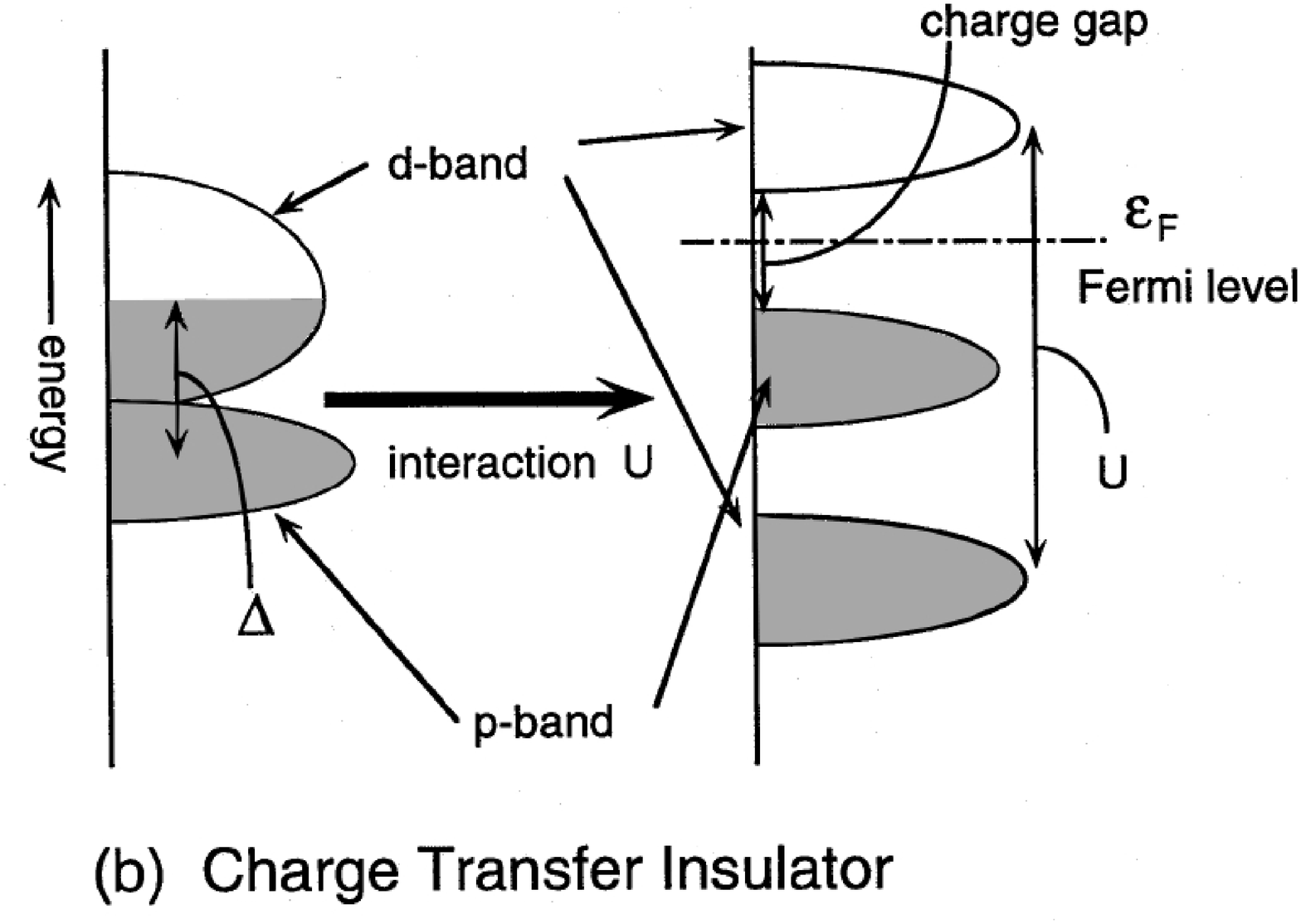}
 \end{minipage}
 \caption{Schematic illustration of the effect of the Coulomb interaction on the energy levels in a Mott-Hubbard (a) and a charge transfer insulator (b). Figure from \cite{mit_review}.}
 \label{insulators}
 \end{figure}

\begin{figure*}[t]
\centering
\begin{minipage}{0.45\linewidth}
\centering
\vspace{15pt}
\includegraphics[width=\linewidth]{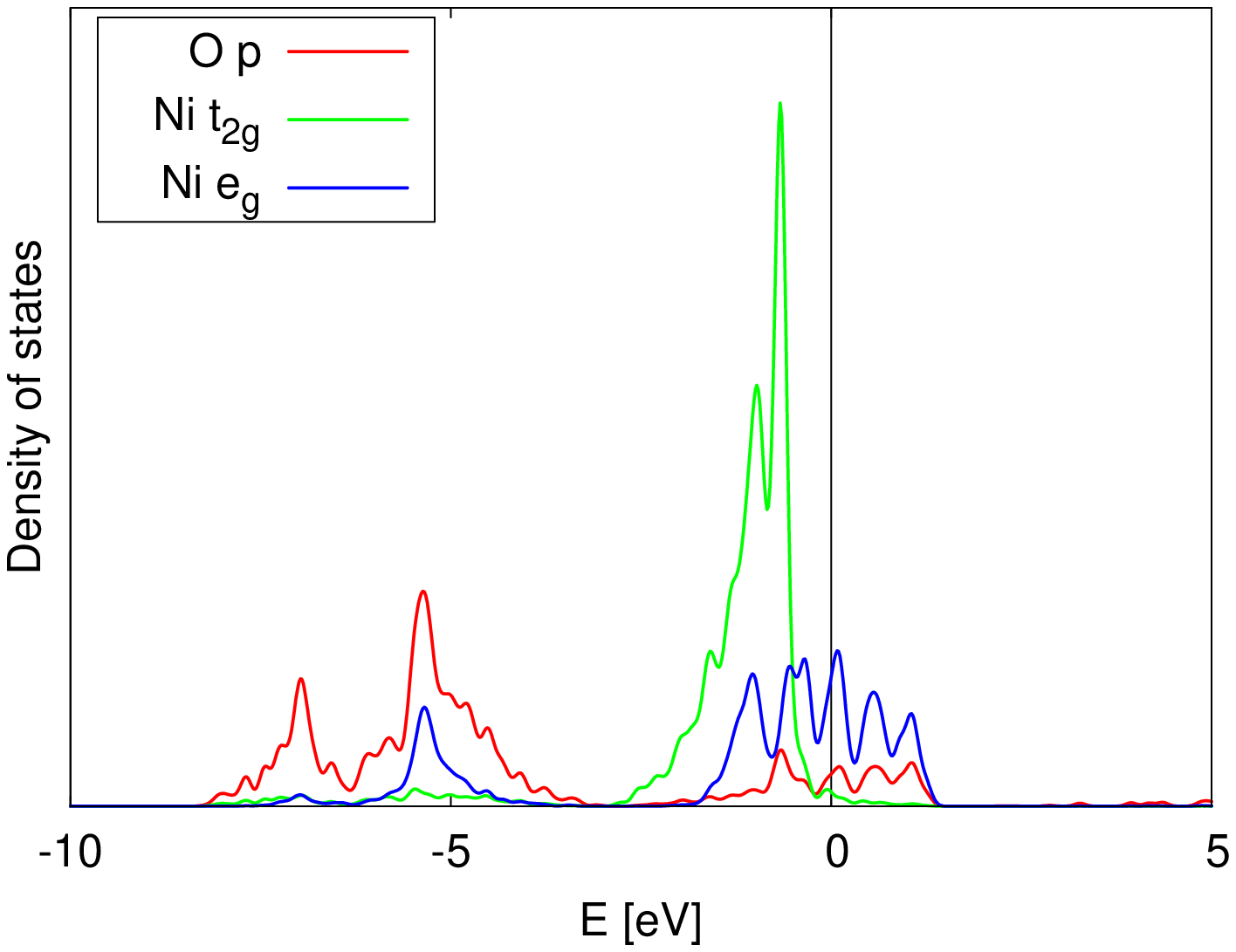}
\end{minipage}
\hspace*{5pt}
\begin{minipage}{0.452\linewidth}
\centering
  \includegraphics[width=\linewidth]{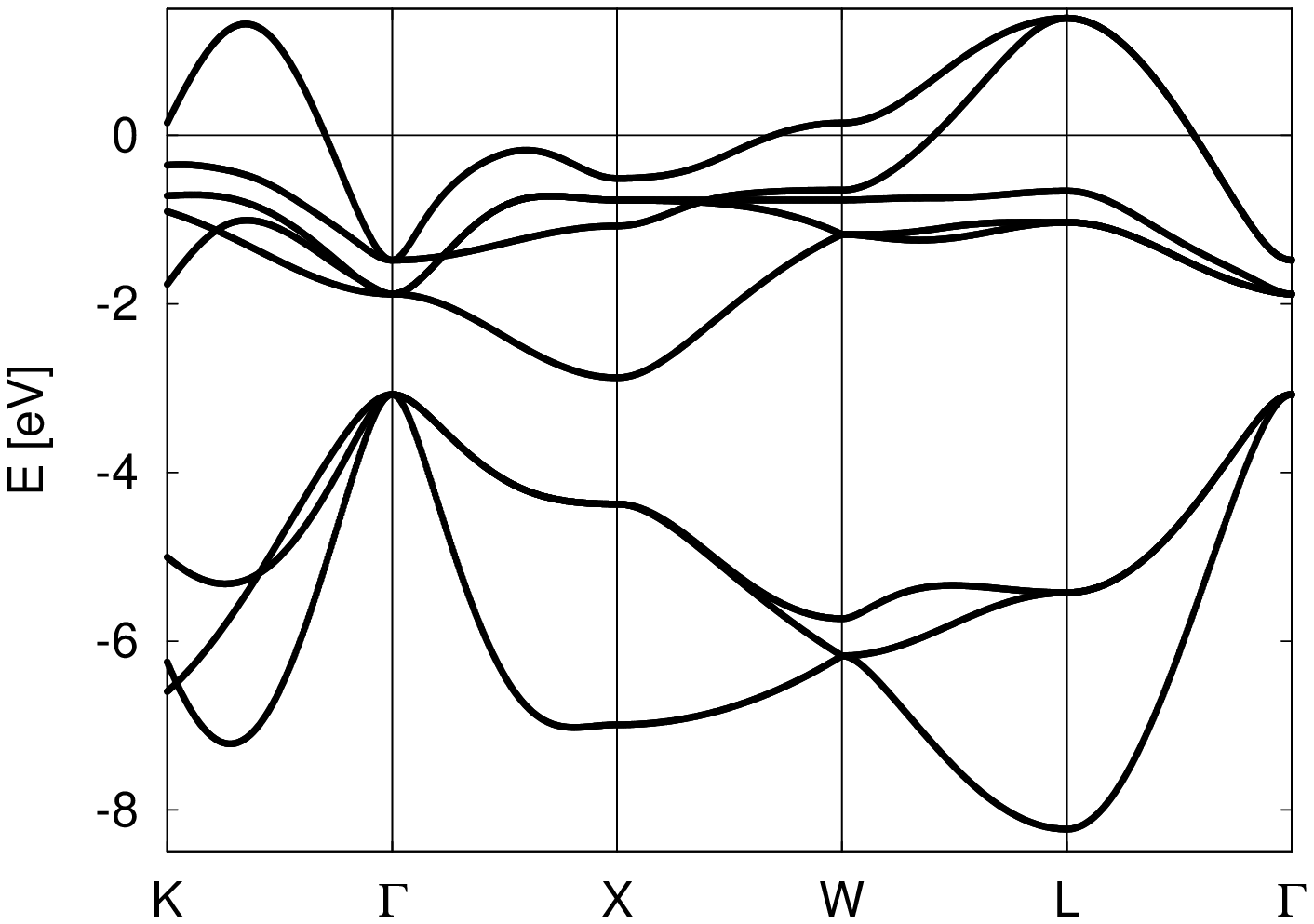}
\end{minipage}
\caption{Density of states (left) and band structure (right) of NiO as obtained by LDA calculations. In the band structure the 5 bands crossing the Fermi level are Ni $3d$ bands, the 3 bands below correspond to oxygen $2p$ states. For further details we refer to the text.}
\label{nio_lda}
\end{figure*}

Nickel oxide is a strongly correlated transition metal oxide that is a prototypic member of the class of charge transfer insulators. According to Zaanen, Sawatzky and Allen transition metal oxides can exhibit a behavior different to the classic Mott-Hubbard picture \cite{zsa}. In a Mott-Hubbard insulator the charge gap opens through splitting of the $d$ band by the Hubbard $U$. In the charge-transfer system the gap typically opens between hybridized ligand $p$ and transition metal $d$ states and the upper Hubbard band corresponding to the $d$ states of the transition metal. Thus, it is not only the Hubbard U, but also the so called charge transfer energy $\Delta=|\varepsilon_d-\varepsilon_p|$ that determines the size of the gap. In the scheme by Zaanen, Sawatzky and Allen materials can be classified by their respective values of $U$ and $\Delta$ \cite{bocquet96}. For $\Delta>U$ the system is a Mott-Hubbard insulator, whereas for $\Delta<U$ it belongs to the charge transfer class. In general, systems with completely filled $d(e_g)$ and partially filled $d(t_{2g})$ shells, like titanates, vanadates and some ruthenates belong to the Mott-Hubbard class. Prominent examples of charge transfer insulators are NiO, MnO, manganites and cuprates. In these systems the $e_g$ shell is partially filled and the $t_{2g}$ shell is fully occupied.

The density of states and the band structure of NiO as obtained by LDA (using the PAW \cite{Bloechl_PAW} based VASP code \cite{kresse_joubert}) are shown in Fig.(\ref{nio_lda}). The band structure shows five Ni $3d$ bands in the energy window $-2.5$eV to $+1.5$eV crossing the Fermi energy and three separated O $2p$ bands below, extending down to $-8$eV. These bands contain 14 electrons in total, 6 occupy the oxygen $p$ bands and the remaining 8 the Ni $d$ bands. In contrast to the LDA prediction NiO is not a metal, on the contrary, experiments revealed a charge gap of about $4 \mathrm{eV}$ \cite{sawatzky_allen}. Additionally, it exhibits antiferromagnetic order below the N{\'e}el temperature of $T_{N}=525\mathrm{K}$. Our computations were carried out in the paramagnetic phase, which is not problematic, since the gap opened by electronic correlations does not depend on whether the system is magnetically ordered. It has been shown in angle-resolved photoemission experiments (ARPES), that passing the N{\'e}el temperature does not qualitatively alter the valence band spectrum \cite{tjernberg}. 

\section{Methodology and Results}

\begin{figure*}[t]
\centering
\subfigure[$\mu_{dc}=21\mathrm{eV}$~,~$\mu=3.0\mathrm{eV}$]{\label{nio_dos_a}\includegraphics[width=0.375\linewidth]{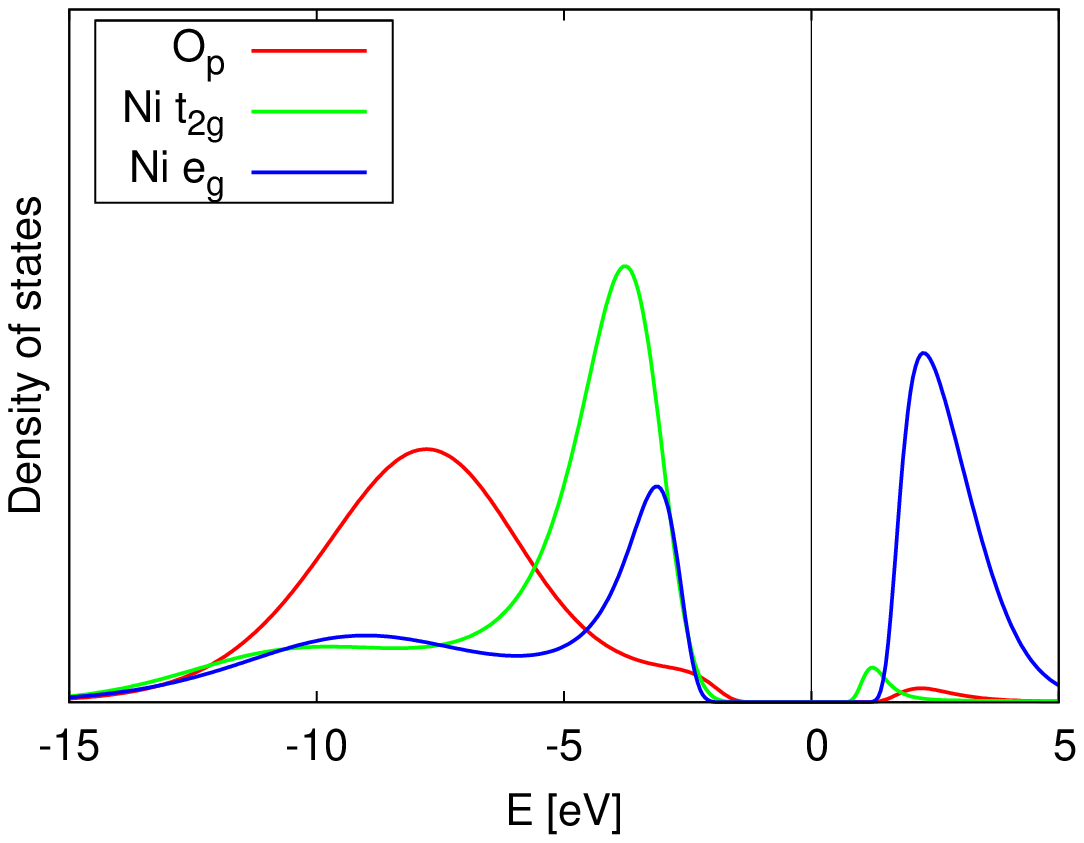}}
\hspace*{50pt}
\subfigure[$\mu_{dc}=25\mathrm{eV}$~,~$\mu=0.5\mathrm{eV}$]{\label{nio_dos_b}\includegraphics[width=0.375\linewidth]{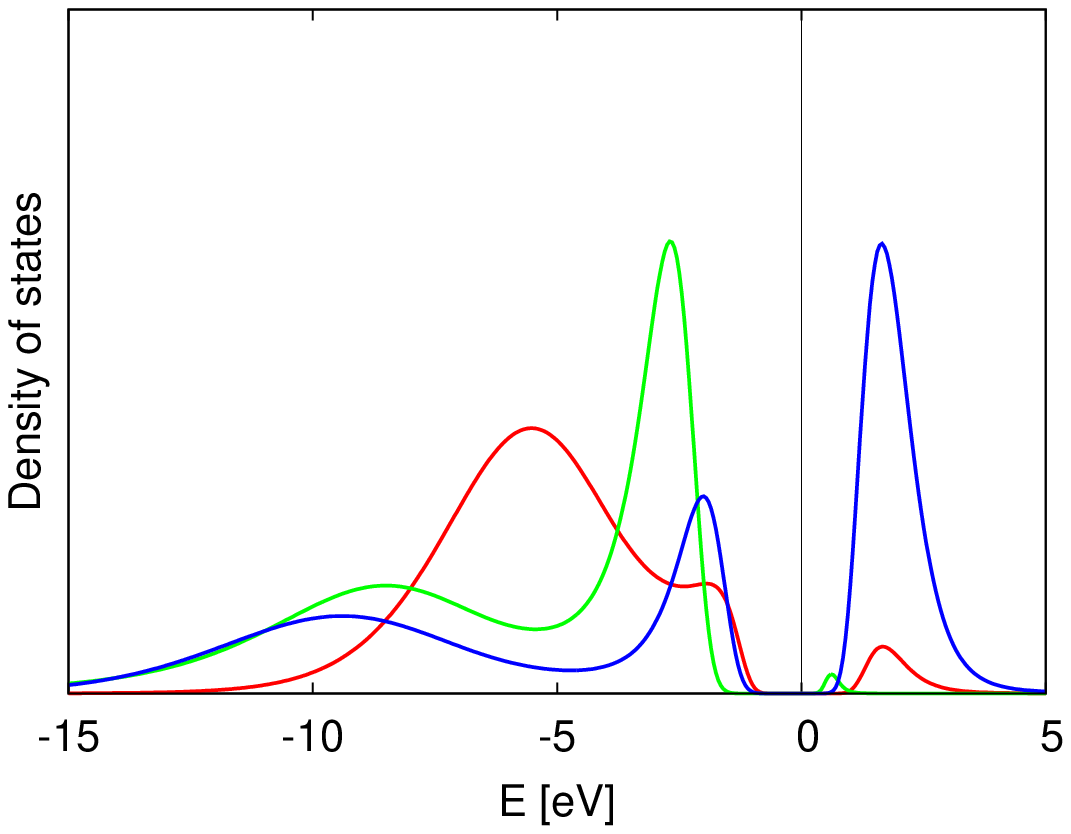}}
\caption{Spectral functions at $\beta=5\mathrm{eV}^{-1}$ for different values of the double counting $\mu_{dc}$ obtained with LDA+DMFT (QMC).}
\label{nio_dos}
\end{figure*}

The model that has to be used for a simulation of NiO is the five band Hubbard model which describes the correlated $3d$ states of Ni. We have calculated the model parameters of such a model in an $ab~initio$ fashion. The local orbitals are represented by Wannier functions, which have been shown recently \cite{anisimov_wannier,pavarini_andersen,lechermann_wannier} to be a very good choice for a basis set, because they form a complete basis of the Hilbert space spanned by Bloch functions and are reminiscent of localized atomic orbitals. Our calculations involved two different flavours of the LDA+DMFT framework: One uses a projection of Bloch states on local orbitals represented by Wannier functions \cite{Amadon08,lechermann_wannier} and a Quantum Monte Carlo (QMC) solver \cite{hirsch_fye}, while the other employs the Linear Order Muffin-Tin Orbital method (LMTO) \cite{andersen_lmto} and a finite-temperature exact diagonalization (ED) solver \cite{caffarel_krauth,capone}.

The effective Wannier Hamiltonian includes the five $3d$ bands of nickel as the correlated subspace and the three $2p$ bands of oxygen as the uncorrelated part.
The inclusion of the $p$ bands is physically motivated since in a charge transfer compound the oxygen bands play an important role in the physics of the system, as was pointed out above. A computation taking into account only the Ni $d$ states is capable of reproducing the insulating behavior and the size of the gap as shown by Ren et al. \cite{ren_nio}. Additionally, the double counting is reduced to a trivial shift in calculations that contain only the Ni $d$ bands, since the full Wannier Hamiltonian belongs to the correlated subspace. The double counting can thus be absorbed into the total chemical potential. However, the physics of the charge transfer insulator cannot be captured without taking into account the ligand $p$ states. 

\medskip

Our calculations were performed at inverse temperature $\beta=5\mathrm{eV}^{-1}$, which corresponds to $2321\mathrm{K}$, using up to $80$ time slices and on the order of $\sim10^6$ Monte Carlo sweeps in the QMC. In the ED fraction of calculations we used a ten site cluster (5 impurity levels and 5 bath levels). The temperature used may appear high, yet it is low enough to give a qualitatively correct description of the physics of the material. Computations at lower temperatures pose no fundamental problem, the amount of LDA+DMFT calculations performed for this study would have made them too expensive though. We have used a Coulomb interaction matrix corresponding to the parameter values $U=8\mathrm{eV}$ and $J=1\mathrm{eV}$.

\begin{figure*}[t]
\centering
\subfigure[$\mu_{dc}=21\mathrm{eV}$]{\label{nio_arpes:a}\includegraphics[width=0.225\linewidth,angle=-90]{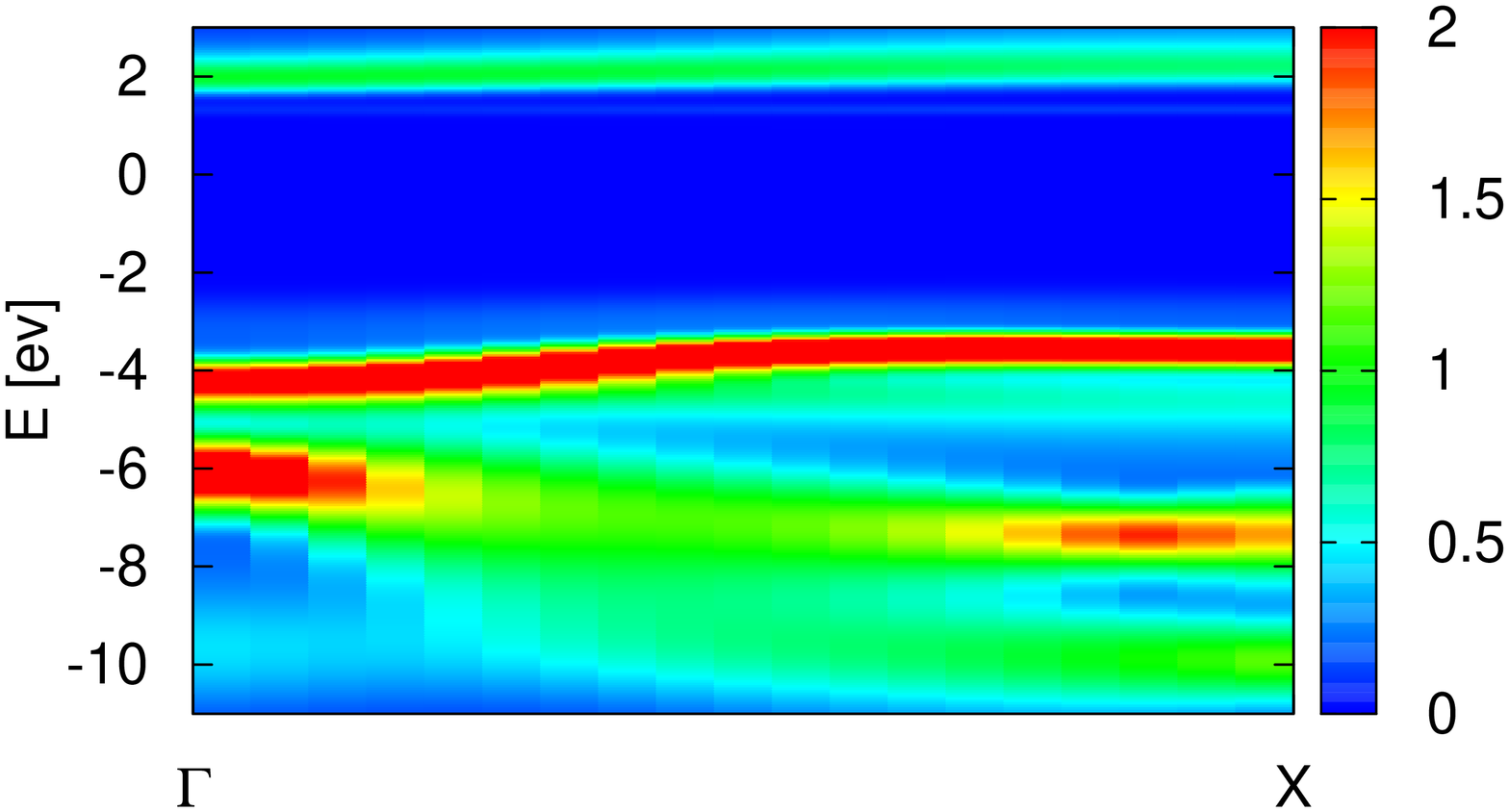}}
\hspace{30pt}
\subfigure[$\mu_{dc}=25\mathrm{eV}$]{\label{nio_arpes:b}\includegraphics[width=0.225\linewidth,angle=-90]{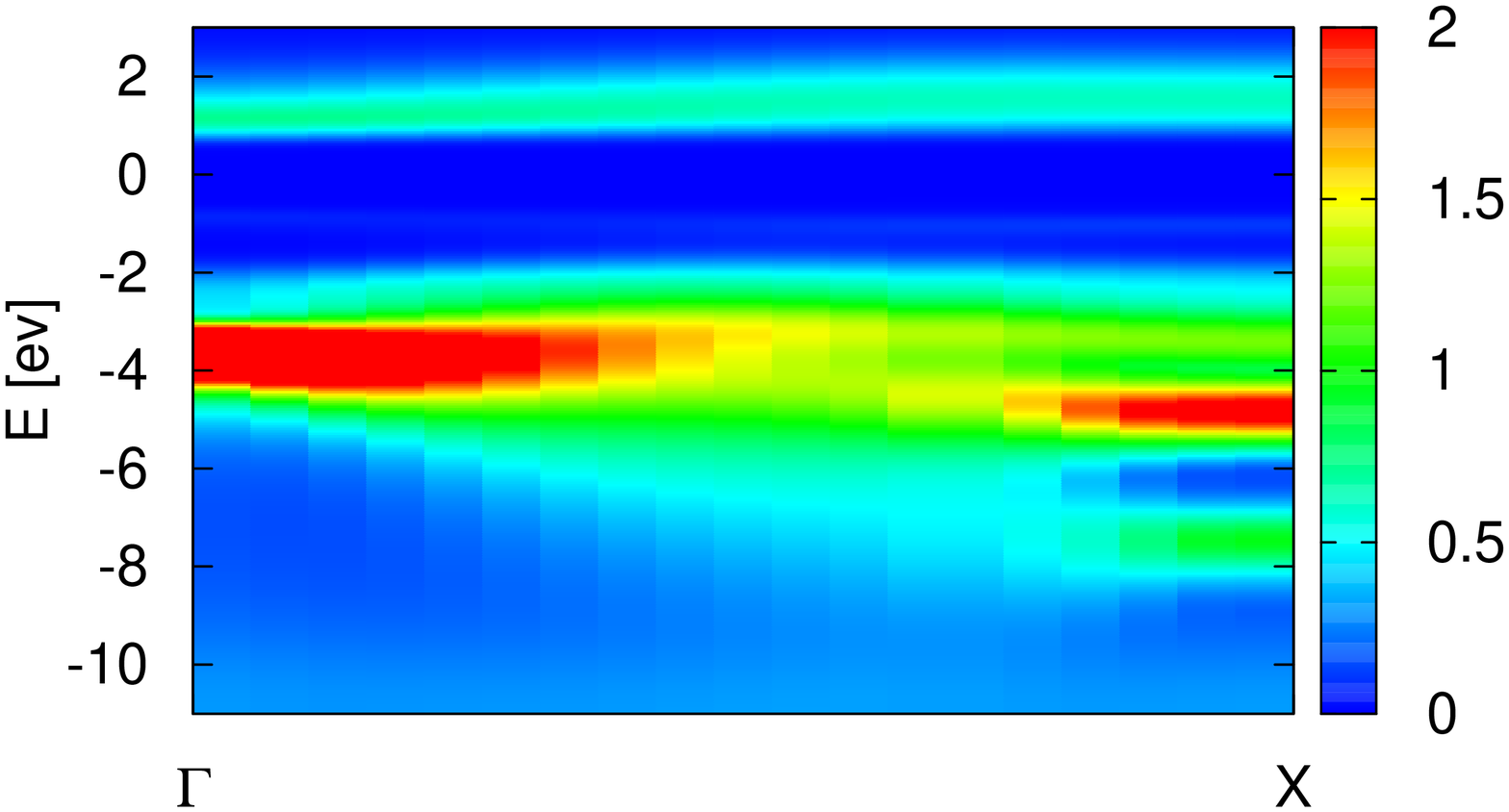}}
\caption{$\mathbf{k}$-resolved spectral functions $A(\mathbf{k},\omega)$ along the line $\Gamma$---$X$ in the Brillouin zone for different values of the double counting $\mu_{dc}$ obtained using LDA+DMFT (QMC).}
\label{nio_arpes}
\end{figure*}

\medskip

The double-counting potential $\mu_{dc}$ defined above is found to have profound impact on the density of states $N_i(\omega)=-\tfrac{1}{\pi}G_i(\omega)$ shown in Fig.(\ref{nio_dos}) and the $\mathbf{k}$-resolved spectral function \newline $A_i(\mathbf{k},\omega)=-\tfrac{1}{\pi}\mathrm{Im}\left(\omega+\mu-\varepsilon_i(\mathbf{k})-\Sigma_i(\omega)\right)^{-1}$ shown along the line $\Gamma$---$X$ in the Brillouin zone in Fig.(\ref{nio_arpes}). The spectral functions were obtained by the maximum entropy method \cite{maxent} from imaginary time Green functions. 
The double-counting potential has been treated here as an adjustable parameter and has been varied between $21\mathrm{eV}$ and $26\mathrm{eV}$. These values already contain the intrinsic shift due to the energy of the particle-hole symmetry in the Hirsch-Fye QMC method that amounts to $34\mathrm{eV}$ with our values of $U$ and $J$. The energy of the particle-hole symmetry is obtained from Eq.(\ref{dc_amf}) with $n^0=\tfrac{1}{2}$. 

The most prominent effects of the double counting on the spectral properties are the shift of the oxygen $p$ bands with respect to the nickel $d$ bands, as well as the variation in gap size. Plainly speaking, the double-counting correction allows for a tuning of the spectral properties from a large gap Mott-Hubbard insulator
to a metal. 
The regime of the charge transfer insulator, the expected physical state of NiO, lies somewhere in between. The experimental spectrum, obtained by x-ray-photoemission (PES) and bremsstrahlung-isochromat-spectroscopy
(BIS) showing both occupied and unoccupied parts, was obtained by e.g. Sawatzky and Allen \cite{sawatzky_allen}. The spectrum recorded at $120 \mathrm{eV}$ is predominantly of Ni $3d$ character, while the $66\mathrm{eV}$ spectrum contains a strong contribution of O $2p$ at about $-4\mathrm{eV}$ \cite{sawatzky_allen,eastman_freeouf}. Additionally, the detailed decomposition of the spectra showed contributions of both O $2p$ and Ni $3d$ at the top of the valence band \cite{sawatzky_allen,eastman_freeouf}. The calculated LDA+DMFT(QMC) spectral function shown in Fig.(\ref{nio_dos}) show basically the two different physical situations of a Mott-Hubbard Fig.(\ref{nio_dos_a}) and a charge-transfer insulator Fig.(\ref{nio_dos_b}) mentioned above.
Both spectral functions were obtained for NiO, by varying the double-counting correction. The characteristic feature of a charge-transfer system, the strongly hybridized ligand $p$ and transition metal $d$ character of the low-energy charge excitations \cite{mit_review,sawatzky_allen}, is only present in the spectrum in Fig.(\ref{nio_dos_b}). The spectrum in Fig.(\ref{nio_dos_a}) is missing this feature almost completely and shows Mott-Hubbard behaviour. This difference underscores the importance of the proper choice for the double-counting correction.

\medskip

Let us now turn to the $\mathbf{k}$-resolved spectral functions shown in Fig.(\ref{nio_arpes}) and compare them with ARPES data \cite{nio_arpes1,nio_arpes2}. The uppermost band in Figs.(\ref{nio_arpes:a}, \ref{nio_arpes:b}) at $\sim2$eV above the Fermi level is a Ni $e_g$ band, while the other bands can be identified with the ones obtained by ARPES. The two lowest lying bands correspond to oxygen $p$ states, the bands above are formed by Ni $d$ states. The characteristic features seen in ARPES, like the broadening of the oxygen bands around the midpoint of the $\Gamma$---$X$ line, are clearly present. The quantitative features, especially the relative band energies can strongly differ, depending on the double counting chosen. 
The bands in Fig.(\ref{nio_arpes:a}) ($\mu_{dc}=21\mathrm{eV}$) show a clear separation between the oxygen and the nickel part at the $\Gamma$-point as well as the $X$-point. At the increased value of the double counting $\mu_{dc}=25\mathrm{eV}$ the oxygen bands are shifted towards the Fermi level, coming to overlap with the Ni $d$ bands at the $\Gamma$ point as in the ARPES data. 
A detailed comparison of the calculated bandstructures with experiments shows that the bands calculated with $\mu_{dc}=25\mathrm{eV}$ agree very well with the experimental data. These calculations reproduce the flat bands at $-4\mathrm{eV}$ and another at about $-2\mathrm{eV}$ becomes more prominently visible at $\mu_{dc}=25\mathrm{eV}$, while it is very faint at $\mu_{dc}=21\mathrm{eV}$. The dispersive bands in the region $-4\mathrm{eV}$ to $-8\mathrm{eV}$ also agree very well with experiment. Our calculations at this value of $\mu_{dc}$ yield very similar results as those obtained by Kune\v{s} et al. \cite{kunes_nio2}. Calculations with other values of the double counting can strongly differ from the experimental data, as shown by the example of $\mu_{dc}=21\mathrm{eV}$ in Fig.(\ref{nio_arpes:a}).

\medskip

\begin{figure*}[t]
 \centering
  \includegraphics[width=0.35\linewidth, angle=-90]{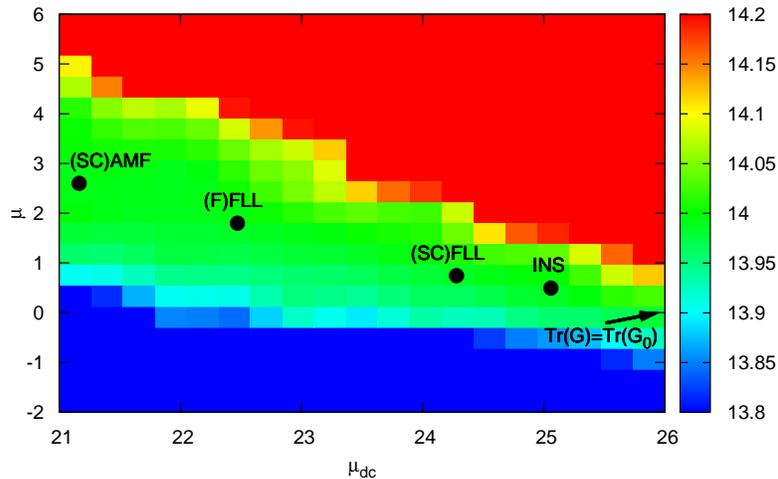}
\caption{Surface created by different combinations of the chemical potential $\mu$ and the double-counting potential $\mu_{dc}$ plotted versus the particle number $N$ obtained with LDA+DMFT (QMC). The particle number has been color coded: the green plateau corresponds to a particle number very close to the desired value of $N=14$, values below are encoded in blue, values above in red. Additionally the results produced by different methods to fix the double counting are indicated. For the AMF and FLL functionals SC or F in parentheses indicates, that the self-consistent occupancies from the DMFT or the formal occupancies have been used respectively. For further details we refer to the text.}
\label{nio_surf}
\end{figure*}

The dimension of the problem of the double counting becomes apparent if the parameter space of the overall chemical potential $\mu$ and the double-counting potential $\mu_{dc}$ versus the total particle number in the system $N$ is examined. The result is shown in Fig.(\ref{nio_surf}) with the particle number color coded. The picture shows that in principle any combination of $\mu$ and $\mu_{dc}$ that yields a point in the green plateau, corresponding to the desired particle number $N=14$ a priori describes the system equivalently good.
The problem that arises here is that conventionally fixing the total chemical potential $\mu$ in the middle of the gap 
still leaves one the freedom of choosing different values for $\mu_{dc}$. An additional condition is required to completely determine the systems position in the $(\mu,\mu_{dc})$ parameter space and thus in the end its spectral properties. As we have argued above this choice is of crucial relevance for the results of the LDA+DMFT simulation and not just an unimportant technicality.

Since other, related approaches, like the LDA+U method, also include a double counting the problem is not new. Over the years different analytic methods to fix $\mu_{dc}$  have been devised. Two prominent examples are the around mean-field (AMF) \cite{ldapu91} approximation and the fully localized or atomic limit (FLL) \cite{amf_fll}. The AMF is based on the conjecture that LDA corresponds to a mean-field solution of the many-body problem, as was argued by Anisimov et al. \cite{ldapu91}. The resulting double-counting potential is
\begin{equation}
\mu_{dc}^{AMF}=\sum_{m^\prime}U_{mm^\prime}n^{0}+
\sum_{m^\prime,m^\prime\neq m}(U_{mm^\prime}-J_{mm^\prime})n^{0},
\label{dc_amf}
\end{equation}
where $n^{0}=\frac{1}{2(2l+1)}\sum_{m,\sigma}n_{m\sigma}$ is the average occupancy. We use the global average and not the spin dependent version proposed in Ref.\cite{amf_fll}, since we were performing paramagnetic calculations in which both spin components are equally occupied. One assumes all orbitals belonging to a certain value of the angular momentum $l$ to be equally occupied and subtracts a corresponding mean-field energy. This is, however, incorrect, since LDA contains the crystal field splitting explicitly and will in general not produce equally occupied orbitals even for weakly correlated systems.   
The result for the case of NiO using self-consistent occupancies from the DMFT loop is shown in Fig.(\ref{nio_surf}) labeled (SC)AMF. The value obtained with the formal occupancies given above ((F)AMF) lies outside of the considered part of the parameter space at $20.4$eV. In both cases the solution corresponds in our case to a Mott-Hubbard insulator as shown in Fig.(\ref{nio_dos_a}). The AMF functional is known to produce unsatisfactory results for strongly correlated systems, which led to the development of another method, the so called FLL.

The FLL functional takes the converse approach to the AMF and begins with the atomic limit. It has been shown, that this new potential can be written as a correction of the AMF solution (\ref{dc_amf}) in the following form \cite{amf_fll}
\[\mu_{dc}^{FLL}=\mu_{dc}^{AMF}+(U-J)(n^0-\tfrac{1}{2}).\]
This addition to the AMF potential has the effect of a shift of the centroid of the level depending on its occupation. An empty level is raised in energy by $\tfrac{1}{2}(U-J)$ and the converse happens to a fully occupied level. The form of the functional is based on the property of the \textit{exact} density functional that the one electron potential should jump discontinuously at integer electron number \cite{ldapu93}, which is not fulfilled in LDA or GGA. Ultimately the FLL leads to a stronger trend towards integer occupancies and localization.
The result of the FLL, as shown in Fig.(\ref{nio_surf}), constitutes a substantial improvement over AMF, yet still produces too low values.
The general problem with analytic expressions like the ones presented is that their scope is limited to certain classes of systems that fulfill the assumptions made in the derivation process. The AMF for example might give good results for weakly correlated systems, but it certainly fails for the strongly correlated ones. The FLL improves the situation for insulators, but it is still based on ad-hoc assumptions. Additionally a certain degree of ambiguity is inherent, since one can compute the corrections using the formal occupancies given above, occupancies obtained from LDA or from the self-consistent DMFT loop.
Other analytical formulas for the double-counting correction have been proposed for the case of NiO, see e.g. the work by Korotin et al. \cite{korotin_wannier} and Kune\v{s} et al. \cite{kunes_nio1}.
Despite giving reasonable resulting spectral functions analytical approaches to the double counting are not optimal.

\medskip

The obvious problems with analytical formulas make conceptually different approaches worth exploring. 
It would certainly be an improvement if the double counting could be found self-consistently along with the chemical potential in the DMFT self-consistency loop.
Since the double counting correction is intrinsically an impurity quantity and not a global quantity (like the chemical potential $\mu$) it would be most desirable to use intrinsic quantities of the impurity like the impurity self-energy or the impurity Green function to fix it. 
One possible ansatz using the impurity self-energy $\Sigma^{imp}_{mm^\prime}$ is to constraint the high energy tails in the real part of the self-energy to sum up to zero
\[\mathrm{Re}\mathrm{Tr}(\Sigma^{imp}_{mm^\prime}(i\omega_N))\overset{!}{=}0.\]
Here, $\omega_N$ is the highest Matsubara frequency included in the computation.
Physically this amounts to the requirement that the shift in the centroid of the impurity orbitals contains no static component. The resulting correction is $\mu_{dc}\sim21.3$eV and thus very close to the (SC)AMF value shown in Fig.(\ref{nio_surf}). 
The result produced is thus reasonable in principle in the sense that it produces an insulating solution. However, the resulting spectrum resembles a Mott-Hubbard system.
Double counting corrections based on the self-energy have been applied successfully to metallic systems, see e.g. \cite{braun}.

\medskip

Another very promising approach, which is in principle based on the Friedel sum rule \cite{hewson}, is to constraint the total charge in the impurity. This approach requires that the electronic charge computed from the local noninteracting Green function and the one computed from the interacting impurity Green function are identical \cite{Amadon08}
\begin{equation}
\mathrm{Tr}~G^{imp}_{mm^\prime}(\beta)\overset{!}{=}\mathrm{Tr}~G^{0,loc}_{mm^\prime}(\beta).
\label{trace}
\end{equation}
Alternatively one can also use the Weiss field $\mathcal{G}_{mm^\prime}$ instead of the local noninteracting Green function in above equation.
Both versions of the method give very similar results and work very well in metallic systems \cite{Amadon08}, since in a metal the total particle number of the system $N$ and of the impurity $n_{imp}$ are both very sensitive to small variations in $\mu$ and $\mu_{dc}$. As NiO has a quite large gap the charge does almost not vary with neither the chemical nor the double-counting potential in the gap. 
The constraint of fixed particle number can thus be fulfilled to a very good approximation in the whole gap region, the criterion (\ref{trace}) essentially breaks down.
Since the gap in NiO is large this method fails and drives the system towards a metallic state at double counting $\mu_{dc}\sim26.5$eV indicated by the arrow pointing out of Fig.(\ref{nio_surf}).


\medskip

\begin{figure*}[t]
\centering
\subfigure[Lattice]{\label{nio_ed_a}\includegraphics[width=0.35\linewidth]{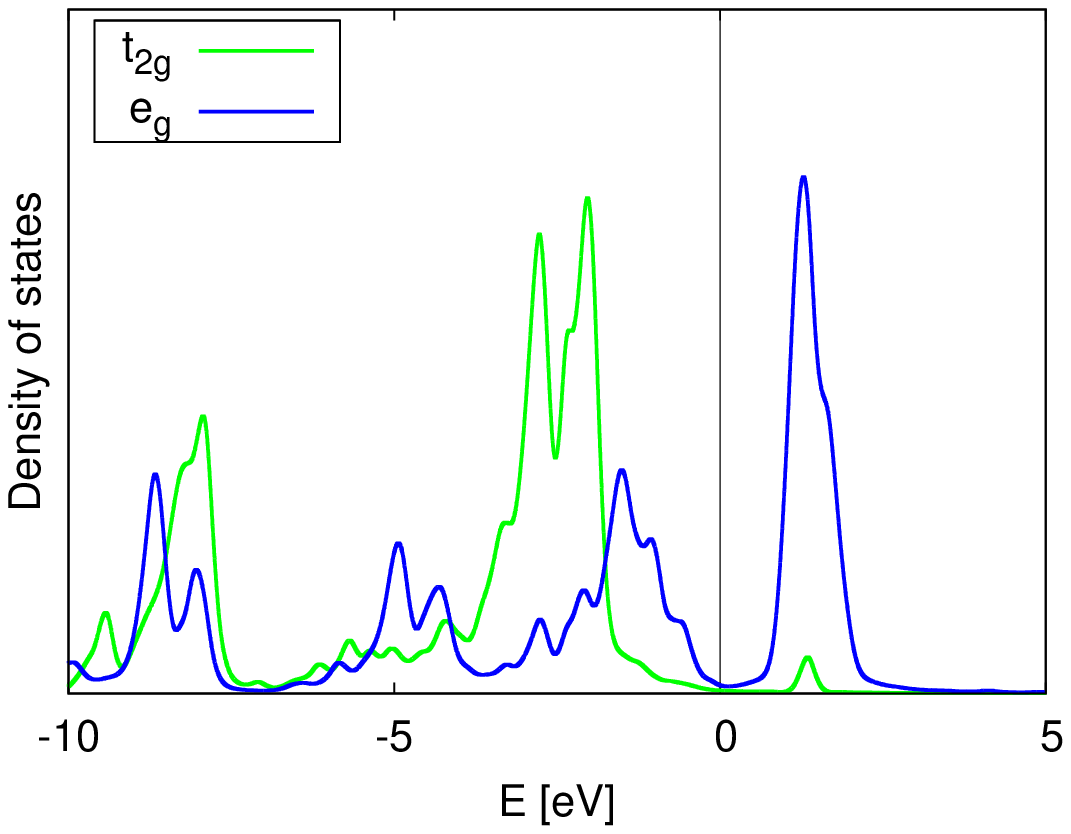}}
\hspace{50pt}
\subfigure[Cluster]{\label{nio_ed_c}\includegraphics[width=0.35\linewidth]{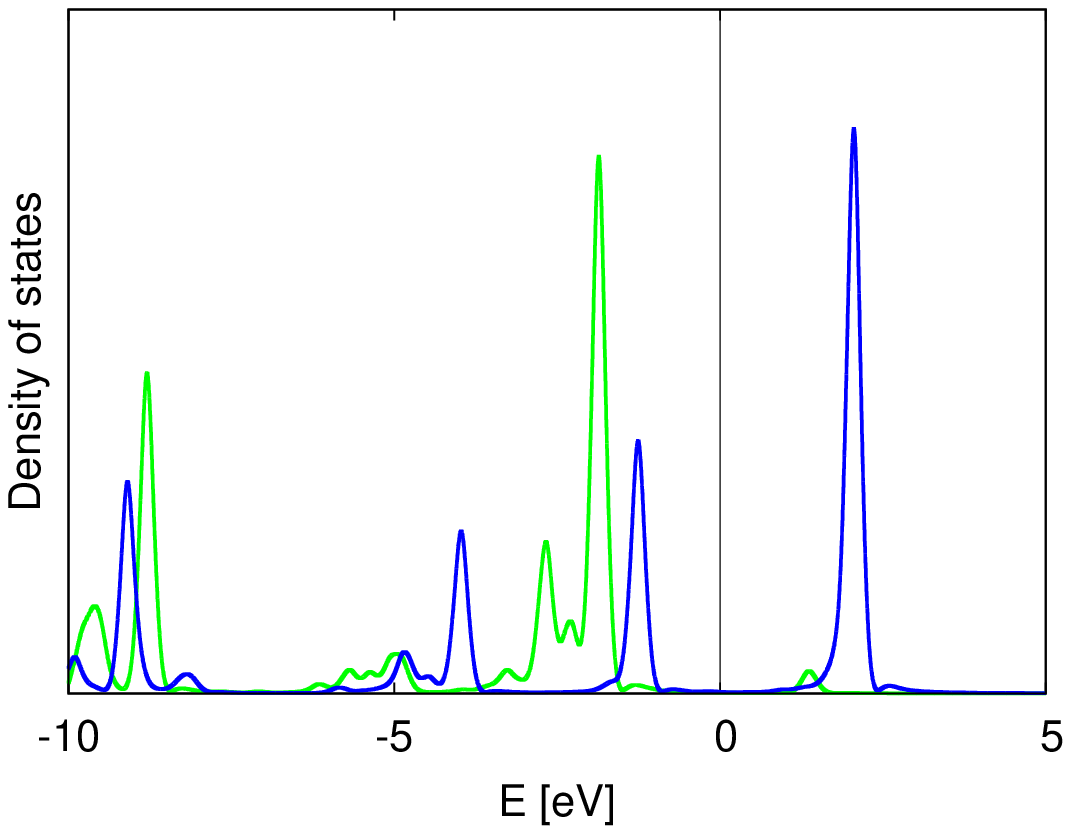}}
\caption{Ni $3d$ spectral functions at $\beta=5\mathrm{eV}^{-1}$ for $\mu_{dc}=25.3\mathrm{eV}$ obtained by LDA+DMFT (ED).}
\label{nio_ed}
\end{figure*}

Since the double-counting corrections that we have explored either fail to reproduce the physics of NiO or are based on analytic arguments that do not exactly apply to the system a different, sound way fixing the value of the double counting for insulating systems is needed. Since the double-counting potential effectively acts like an impurity chemical potential we propose to find the value at which it lies in the middle of the gap of the impurity spectral function where the occupation of the impurity is about $n_{imp}\approx8$ particles. This part of the calculations was done using the exact diagonalization impurity solver (see above), which is much faster and uses the full Coulomb interaction matrix including spin-flip and pair-hopping terms. Additionally it does not suffer from statistical errors and directly provides data on the real axis. We used a 10 site cluster with 5 impurity levels plus 5 bath levels and fit the bath Green function via the level energies and hopping parameters \cite{capone}. An explicit scan of the parameter space revealed that the proper value for the double-counting correction is $\mu_{dc}\sim25\mathrm{eV}$, basically the same value found above by inspection and comparison of spectral features to experimental data. It is indicated as INS in Fig.(\ref{nio_surf}). The corresponding lattice and cluster spectral functions are shown in Fig.(\ref{nio_ed}). The proposed criterion thus produces a double-counting correction that reproduces the spectral features of the valence in accord with photoemmission measurements and does not contain ad-hoc assumptions about the system.  

\section{Summary and Conclusions}
In summary, our study has shown that the double-counting correction in the LDA+DMFT formalism has to be very carefully assessed when performing calculations with a correlated and uncorrelated part in the Hamiltonian. We have examined the influence of the double-counting potential on the spectral properties using the example of NiO. Different tracks in the search for a sound double counting were explored. A well defined analytical expression for the double-counting potential $\mu_{dc}$ probably cannot be formulated in the context of LDA+DMFT. 
Thus, one has to resort to numerical criteria to fix the value of the double-counting correction.
For metals the self consistency criterion based on the charge Eq.(\ref{trace}) works very reliably. It is, however, not applicable to insulating systems. In such a case we proposed to fix the value of the double-counting potential by setting it in the middle of the gap of the impurity  spectral function. This criterion led to spectral properties in good agreement with experiments.
Thus, one has to resort to self-consistent numerical approaches to fix the double-counting correction properly.
Further work, especially the examination of other systems will show if the proposed methodology can be reliably applied to predict the electronic structure of correlated electron systems by LDA+DMFT calculations.


\begin{thebibliography}{31}
\expandafter\ifx\csname natexlab\endcsname\relax\def\natexlab#1{#1}\fi
\expandafter\ifx\csname bibnamefont\endcsname\relax
  \def\bibnamefont#1{#1}\fi
\expandafter\ifx\csname bibfnamefont\endcsname\relax
  \def\bibfnamefont#1{#1}\fi
\expandafter\ifx\csname citenamefont\endcsname\relax
  \def\citenamefont#1{#1}\fi
\expandafter\ifx\csname url\endcsname\relax
  \def\url#1{\texttt{#1}}\fi
\expandafter\ifx\csname urlprefix\endcsname\relax\def\urlprefix{URL }\fi
\providecommand{\bibinfo}[2]{#2}
\providecommand{\eprint}[2][]{\url{#2}}

\bibitem[{\citenamefont{Georges et~al.}(1996)\citenamefont{Georges, Kotliar,
  Krauth, and Rozenberg}}]{dmft_review96}
\bibinfo{author}{\bibfnamefont{A.}~\bibnamefont{Georges}},
  \bibinfo{author}{\bibfnamefont{G.}~\bibnamefont{Kotliar}},
  \bibinfo{author}{\bibfnamefont{W.}~\bibnamefont{Krauth}}, \bibnamefont{and}
  \bibinfo{author}{\bibfnamefont{M.~J.} \bibnamefont{Rozenberg}},
  \bibinfo{journal}{Rev. Mod. Phys.} \textbf{\bibinfo{volume}{68}},
  \bibinfo{pages}{13} (\bibinfo{year}{1996}).

\bibitem[{\citenamefont{Anisimov et~al.}(1997)\citenamefont{Anisimov,
  Poteryaev, Korotin, Anokhin, and Kotliar}}]{anisimov_dmft}
\bibinfo{author}{\bibfnamefont{V.~I.} \bibnamefont{Anisimov}},
  \bibinfo{author}{\bibfnamefont{A.~I.} \bibnamefont{Poteryaev}},
  \bibinfo{author}{\bibfnamefont{M.~A.} \bibnamefont{Korotin}},
  \bibinfo{author}{\bibfnamefont{A.~O.} \bibnamefont{Anokhin}},
  \bibnamefont{and} \bibinfo{author}{\bibfnamefont{G.}~\bibnamefont{Kotliar}},
  \bibinfo{journal}{Journal of Physics: Condensed Matter}
  \textbf{\bibinfo{volume}{9}}, \bibinfo{pages}{7359} (\bibinfo{year}{1997}).

\bibitem[{\citenamefont{Lichtenstein and Katsnelson}(1998)}]{lda++}
\bibinfo{author}{\bibfnamefont{A.~I.} \bibnamefont{Lichtenstein}}
  \bibnamefont{and} \bibinfo{author}{\bibfnamefont{M.~I.}
  \bibnamefont{Katsnelson}}, \bibinfo{journal}{Phys. Rev. B}
  \textbf{\bibinfo{volume}{57}}, \bibinfo{pages}{6884} (\bibinfo{year}{1998}).

\bibitem[{\citenamefont{Ren et~al.}(2006)\citenamefont{Ren, Leonov, Keller,
  Kollar, Nekrasov, and Vollhardt}}]{ren_nio}
\bibinfo{author}{\bibfnamefont{X.}~\bibnamefont{Ren}},
  \bibinfo{author}{\bibfnamefont{I.}~\bibnamefont{Leonov}},
  \bibinfo{author}{\bibfnamefont{G.}~\bibnamefont{Keller}},
  \bibinfo{author}{\bibfnamefont{M.}~\bibnamefont{Kollar}},
  \bibinfo{author}{\bibfnamefont{I.}~\bibnamefont{Nekrasov}}, \bibnamefont{and}
  \bibinfo{author}{\bibfnamefont{D.}~\bibnamefont{Vollhardt}},
  \bibinfo{journal}{Phys. Rev. B} \textbf{\bibinfo{volume}{74}},
  \bibinfo{eid}{195114} (pages~\bibinfo{numpages}{8}) (\bibinfo{year}{2006}).

\bibitem[{\citenamefont{Kune\v{s}
  et~al.}(2007{\natexlab{a}})\citenamefont{Kune\v{s}, Anisimov, Lukoyanov, and
  Vollhardt}}]{kunes_nio1}
\bibinfo{author}{\bibfnamefont{J.}~\bibnamefont{Kune\v{s}}},
  \bibinfo{author}{\bibfnamefont{V.~I.} \bibnamefont{Anisimov}},
  \bibinfo{author}{\bibfnamefont{A.~V.} \bibnamefont{Lukoyanov}},
  \bibnamefont{and}
  \bibinfo{author}{\bibfnamefont{D.}~\bibnamefont{Vollhardt}},
  \bibinfo{journal}{Phys. Rev. B} \textbf{\bibinfo{volume}{75}},
  \bibinfo{eid}{165115} (pages~\bibinfo{numpages}{4})
  (\bibinfo{year}{2007}{\natexlab{a}}).

\bibitem[{\citenamefont{Kune\v{s}
  et~al.}(2007{\natexlab{b}})\citenamefont{Kune\v{s}, Anisimov, Skornyakov,
  Lukoyanov, and Vollhardt}}]{kunes_nio2}
\bibinfo{author}{\bibfnamefont{J.}~\bibnamefont{Kune\v{s}}},
  \bibinfo{author}{\bibfnamefont{V.~I.} \bibnamefont{Anisimov}},
  \bibinfo{author}{\bibfnamefont{S.~L.} \bibnamefont{Skornyakov}},
  \bibinfo{author}{\bibfnamefont{A.~V.} \bibnamefont{Lukoyanov}},
  \bibnamefont{and}
  \bibinfo{author}{\bibfnamefont{D.}~\bibnamefont{Vollhardt}},
  \bibinfo{journal}{Phys. Rev. Lett.} \textbf{\bibinfo{volume}{99}},
  \bibinfo{eid}{156404} (pages~\bibinfo{numpages}{4})
  (\bibinfo{year}{2007}{\natexlab{b}}).

\bibitem[{\citenamefont{Korotin et~al.}(2008)\citenamefont{Korotin,
  Kozhevnikov, Skornyakov, Leonov, Binggeli, Anisimov, and
  Trimarchi}}]{korotin_wannier}
\bibinfo{author}{\bibfnamefont{D.}~\bibnamefont{Korotin}},
  \bibinfo{author}{\bibfnamefont{A.~V.} \bibnamefont{Kozhevnikov}},
  \bibinfo{author}{\bibfnamefont{S.~L.} \bibnamefont{Skornyakov}},
  \bibinfo{author}{\bibfnamefont{I.}~\bibnamefont{Leonov}},
  \bibinfo{author}{\bibfnamefont{N.}~\bibnamefont{Binggeli}},
  \bibinfo{author}{\bibfnamefont{V.~I.} \bibnamefont{Anisimov}},
  \bibnamefont{and}
  \bibinfo{author}{\bibfnamefont{G.}~\bibnamefont{Trimarchi}},
  \bibinfo{journal}{The European Physical Journal B}
  \textbf{\bibinfo{volume}{65}}, \bibinfo{pages}{91} (\bibinfo{year}{2008}).

\bibitem[{\citenamefont{Imada et~al.}(1998)\citenamefont{Imada, Fujimori, and
  Tokura}}]{mit_review}
\bibinfo{author}{\bibfnamefont{M.}~\bibnamefont{Imada}},
  \bibinfo{author}{\bibfnamefont{A.}~\bibnamefont{Fujimori}}, \bibnamefont{and}
  \bibinfo{author}{\bibfnamefont{Y.}~\bibnamefont{Tokura}},
  \bibinfo{journal}{Rev. Mod. Phys.} \textbf{\bibinfo{volume}{70}},
  \bibinfo{pages}{1039} (\bibinfo{year}{1998}).

\bibitem[{\citenamefont{Zaanen et~al.}(1985)\citenamefont{Zaanen, Sawatzky, and
  Allen}}]{zsa}
\bibinfo{author}{\bibfnamefont{J.}~\bibnamefont{Zaanen}},
  \bibinfo{author}{\bibfnamefont{G.~A.} \bibnamefont{Sawatzky}},
  \bibnamefont{and} \bibinfo{author}{\bibfnamefont{J.~W.} \bibnamefont{Allen}},
  \bibinfo{journal}{Phys. Rev. Lett.} \textbf{\bibinfo{volume}{55}},
  \bibinfo{pages}{418} (\bibinfo{year}{1985}).

\bibitem[{\citenamefont{Bocquet et~al.}(1996)\citenamefont{Bocquet, Mizokawa,
  Morikawa, Fujimori, Barman, Maiti, Sarma, Tokura, and Onoda}}]{bocquet96}
\bibinfo{author}{\bibfnamefont{A.~E.} \bibnamefont{Bocquet}},
  \bibinfo{author}{\bibfnamefont{T.}~\bibnamefont{Mizokawa}},
  \bibinfo{author}{\bibfnamefont{K.}~\bibnamefont{Morikawa}},
  \bibinfo{author}{\bibfnamefont{A.}~\bibnamefont{Fujimori}},
  \bibinfo{author}{\bibfnamefont{S.~R.} \bibnamefont{Barman}},
  \bibinfo{author}{\bibfnamefont{K.}~\bibnamefont{Maiti}},
  \bibinfo{author}{\bibfnamefont{D.~D.} \bibnamefont{Sarma}},
  \bibinfo{author}{\bibfnamefont{Y.}~\bibnamefont{Tokura}}, \bibnamefont{and}
  \bibinfo{author}{\bibfnamefont{M.}~\bibnamefont{Onoda}},
  \bibinfo{journal}{Phys. Rev. B} \textbf{\bibinfo{volume}{53}},
  \bibinfo{pages}{1161} (\bibinfo{year}{1996}).

\bibitem[{\citenamefont{Bl\"ochl}(1994)}]{Bloechl_PAW}
\bibinfo{author}{\bibfnamefont{P.~E.} \bibnamefont{Bl\"ochl}},
  \bibinfo{journal}{Phys. Rev. B} \textbf{\bibinfo{volume}{50}},
  \bibinfo{pages}{17953} (\bibinfo{year}{1994}).

\bibitem[{\citenamefont{Kresse and Joubert}(1999)}]{kresse_joubert}
\bibinfo{author}{\bibfnamefont{G.}~\bibnamefont{Kresse}} \bibnamefont{and}
  \bibinfo{author}{\bibfnamefont{D.}~\bibnamefont{Joubert}},
  \bibinfo{journal}{Phys. Rev. B} \textbf{\bibinfo{volume}{59}},
  \bibinfo{pages}{1758} (\bibinfo{year}{1999}).

\bibitem[{\citenamefont{Sawatzky and Allen}(1984)}]{sawatzky_allen}
\bibinfo{author}{\bibfnamefont{G.~A.} \bibnamefont{Sawatzky}} \bibnamefont{and}
  \bibinfo{author}{\bibfnamefont{J.~W.} \bibnamefont{Allen}},
  \bibinfo{journal}{Phys. Rev. Lett.} \textbf{\bibinfo{volume}{53}},
  \bibinfo{pages}{2339} (\bibinfo{year}{1984}).

\bibitem[{\citenamefont{Tjernberg et~al.}(1996)\citenamefont{Tjernberg,
  S\"oderholm, Chiaia, Girard, Karlsson, Nyl\'en, and Lindau}}]{tjernberg}
\bibinfo{author}{\bibfnamefont{O.}~\bibnamefont{Tjernberg}},
  \bibinfo{author}{\bibfnamefont{S.}~\bibnamefont{S\"oderholm}},
  \bibinfo{author}{\bibfnamefont{G.}~\bibnamefont{Chiaia}},
  \bibinfo{author}{\bibfnamefont{R.}~\bibnamefont{Girard}},
  \bibinfo{author}{\bibfnamefont{U.~O.} \bibnamefont{Karlsson}},
  \bibinfo{author}{\bibfnamefont{H.}~\bibnamefont{Nyl\'en}}, \bibnamefont{and}
  \bibinfo{author}{\bibfnamefont{I.}~\bibnamefont{Lindau}},
  \bibinfo{journal}{Phys. Rev. B} \textbf{\bibinfo{volume}{54}},
  \bibinfo{pages}{10245} (\bibinfo{year}{1996}).

\bibitem[{\citenamefont{Anisimov et~al.}(2005)\citenamefont{Anisimov, Kondakov,
  Kozhevnikov, Nekrasov, Pchelkina, Allen, Mo, Kim, Metcalf, Suga
  et~al.}}]{anisimov_wannier}
\bibinfo{author}{\bibfnamefont{V.~I.} \bibnamefont{Anisimov}},
  \bibinfo{author}{\bibfnamefont{D.~E.} \bibnamefont{Kondakov}},
  \bibinfo{author}{\bibfnamefont{A.~V.} \bibnamefont{Kozhevnikov}},
  \bibinfo{author}{\bibfnamefont{I.~A.} \bibnamefont{Nekrasov}},
  \bibinfo{author}{\bibfnamefont{Z.~V.} \bibnamefont{Pchelkina}},
  \bibinfo{author}{\bibfnamefont{J.~W.} \bibnamefont{Allen}},
  \bibinfo{author}{\bibfnamefont{S.-K.} \bibnamefont{Mo}},
  \bibinfo{author}{\bibfnamefont{H.-D.} \bibnamefont{Kim}},
  \bibinfo{author}{\bibfnamefont{P.}~\bibnamefont{Metcalf}},
  \bibinfo{author}{\bibfnamefont{S.}~\bibnamefont{Suga}}, \bibnamefont{et~al.},
  \bibinfo{journal}{Phys. Rev. B} \textbf{\bibinfo{volume}{71}},
  \bibinfo{eid}{125119} (\bibinfo{year}{2005}).

\bibitem[{\citenamefont{Pavarini et~al.}(2004)\citenamefont{Pavarini, Biermann,
  Poteryaev, Lichtenstein, Georges, and Andersen}}]{pavarini_andersen}
\bibinfo{author}{\bibfnamefont{E.}~\bibnamefont{Pavarini}},
  \bibinfo{author}{\bibfnamefont{S.}~\bibnamefont{Biermann}},
  \bibinfo{author}{\bibfnamefont{A.}~\bibnamefont{Poteryaev}},
  \bibinfo{author}{\bibfnamefont{A.~I.} \bibnamefont{Lichtenstein}},
  \bibinfo{author}{\bibfnamefont{A.}~\bibnamefont{Georges}}, \bibnamefont{and}
  \bibinfo{author}{\bibfnamefont{O.~K.} \bibnamefont{Andersen}},
  \bibinfo{journal}{Phys. Rev. Lett.} \textbf{\bibinfo{volume}{92}},
  \bibinfo{pages}{176403} (\bibinfo{year}{2004}).

\bibitem[{\citenamefont{Lechermann et~al.}(2006)\citenamefont{Lechermann,
  Georges, Poteryaev, Biermann, Posternak, Yamasaki, and
  Andersen}}]{lechermann_wannier}
\bibinfo{author}{\bibfnamefont{F.}~\bibnamefont{Lechermann}},
  \bibinfo{author}{\bibfnamefont{A.}~\bibnamefont{Georges}},
  \bibinfo{author}{\bibfnamefont{A.}~\bibnamefont{Poteryaev}},
  \bibinfo{author}{\bibfnamefont{S.}~\bibnamefont{Biermann}},
  \bibinfo{author}{\bibfnamefont{M.}~\bibnamefont{Posternak}},
  \bibinfo{author}{\bibfnamefont{A.}~\bibnamefont{Yamasaki}}, \bibnamefont{and}
  \bibinfo{author}{\bibfnamefont{O.~K.} \bibnamefont{Andersen}},
  \bibinfo{journal}{Phys. Rev. B} \textbf{\bibinfo{volume}{74}},
  \bibinfo{eid}{125120} (\bibinfo{year}{2006}).

\bibitem[{\citenamefont{Amadon et~al.}(2008)\citenamefont{Amadon, Lechermann,
  Georges, Jollet, Wehling, and Lichtenstein}}]{Amadon08}
\bibinfo{author}{\bibfnamefont{B.}~\bibnamefont{Amadon}},
  \bibinfo{author}{\bibfnamefont{F.}~\bibnamefont{Lechermann}},
  \bibinfo{author}{\bibfnamefont{A.}~\bibnamefont{Georges}},
  \bibinfo{author}{\bibfnamefont{F.}~\bibnamefont{Jollet}},
  \bibinfo{author}{\bibfnamefont{T.~O.} \bibnamefont{Wehling}},
  \bibnamefont{and} \bibinfo{author}{\bibfnamefont{A.~I.}
  \bibnamefont{Lichtenstein}}, \bibinfo{journal}{Phys. Rev. B}
  \textbf{\bibinfo{volume}{77}}, \bibinfo{eid}{205112} (\bibinfo{year}{2008}).

\bibitem[{\citenamefont{Hirsch and Fye}(1986)}]{hirsch_fye}
\bibinfo{author}{\bibfnamefont{J.~E.} \bibnamefont{Hirsch}} \bibnamefont{and}
  \bibinfo{author}{\bibfnamefont{R.~M.} \bibnamefont{Fye}},
  \bibinfo{journal}{Phys. Rev. Lett.} \textbf{\bibinfo{volume}{56}},
  \bibinfo{pages}{2521} (\bibinfo{year}{1986}).

\bibitem[{\citenamefont{Andersen}(1975)}]{andersen_lmto}
\bibinfo{author}{\bibfnamefont{O.~K.} \bibnamefont{Andersen}},
  \bibinfo{journal}{Phys. Rev. B} \textbf{\bibinfo{volume}{12}},
  \bibinfo{pages}{3060} (\bibinfo{year}{1975}).

\bibitem[{\citenamefont{Caffarel and Krauth}(1994)}]{caffarel_krauth}
\bibinfo{author}{\bibfnamefont{M.}~\bibnamefont{Caffarel}} \bibnamefont{and}
  \bibinfo{author}{\bibfnamefont{W.}~\bibnamefont{Krauth}},
  \bibinfo{journal}{Phys. Rev. Lett.} \textbf{\bibinfo{volume}{72}},
  \bibinfo{pages}{1545} (\bibinfo{year}{1994}).

\bibitem[{\citenamefont{Capone et~al.}(2007)\citenamefont{Capone, de\char39{}
  Medici, and Georges}}]{capone}
\bibinfo{author}{\bibfnamefont{M.}~\bibnamefont{Capone}},
  \bibinfo{author}{\bibfnamefont{L.}~\bibnamefont{de\char39{} Medici}},
  \bibnamefont{and} \bibinfo{author}{\bibfnamefont{A.}~\bibnamefont{Georges}},
  \bibinfo{journal}{Phys. Rev. B} \textbf{\bibinfo{volume}{76}},
  \bibinfo{pages}{245116} (\bibinfo{year}{2007}).

\bibitem[{\citenamefont{Jarrell and Gubernatis}(1996)}]{maxent}
\bibinfo{author}{\bibfnamefont{M.}~\bibnamefont{Jarrell}} \bibnamefont{and}
  \bibinfo{author}{\bibfnamefont{J.~E.} \bibnamefont{Gubernatis}},
  \bibinfo{journal}{Physics Reports} \textbf{\bibinfo{volume}{269}},
  \bibinfo{pages}{133 } (\bibinfo{year}{1996}).

\bibitem[{\citenamefont{Eastman and Freeouf}(1975)}]{eastman_freeouf}
\bibinfo{author}{\bibfnamefont{D.~E.} \bibnamefont{Eastman}} \bibnamefont{and}
  \bibinfo{author}{\bibfnamefont{J.~L.} \bibnamefont{Freeouf}},
  \bibinfo{journal}{Phys. Rev. Lett.} \textbf{\bibinfo{volume}{34}},
  \bibinfo{pages}{395} (\bibinfo{year}{1975}).

\bibitem[{\citenamefont{Shen et~al.}(1990)\citenamefont{Shen, Shih, Jepsen,
  Spicer, Lindau, and Allen}}]{nio_arpes1}
\bibinfo{author}{\bibfnamefont{Z.-X.} \bibnamefont{Shen}},
  \bibinfo{author}{\bibfnamefont{C.~K.} \bibnamefont{Shih}},
  \bibinfo{author}{\bibfnamefont{O.}~\bibnamefont{Jepsen}},
  \bibinfo{author}{\bibfnamefont{W.~E.} \bibnamefont{Spicer}},
  \bibinfo{author}{\bibfnamefont{I.}~\bibnamefont{Lindau}}, \bibnamefont{and}
  \bibinfo{author}{\bibfnamefont{J.~W.} \bibnamefont{Allen}},
  \bibinfo{journal}{Phys. Rev. Lett.} \textbf{\bibinfo{volume}{64}},
  \bibinfo{pages}{2442} (\bibinfo{year}{1990}).

\bibitem[{\citenamefont{Shen et~al.}(1991)\citenamefont{Shen, List, Dessau,
  Wells, Jepsen, Arko, Barttlet, Shih, Parmigiani, Huang et~al.}}]{nio_arpes2}
\bibinfo{author}{\bibfnamefont{Z.-X.} \bibnamefont{Shen}},
  \bibinfo{author}{\bibfnamefont{R.~S.} \bibnamefont{List}},
  \bibinfo{author}{\bibfnamefont{D.~S.} \bibnamefont{Dessau}},
  \bibinfo{author}{\bibfnamefont{B.~O.} \bibnamefont{Wells}},
  \bibinfo{author}{\bibfnamefont{O.}~\bibnamefont{Jepsen}},
  \bibinfo{author}{\bibfnamefont{A.~J.} \bibnamefont{Arko}},
  \bibinfo{author}{\bibfnamefont{R.}~\bibnamefont{Barttlet}},
  \bibinfo{author}{\bibfnamefont{C.~K.} \bibnamefont{Shih}},
  \bibinfo{author}{\bibfnamefont{F.}~\bibnamefont{Parmigiani}},
  \bibinfo{author}{\bibfnamefont{J.~C.} \bibnamefont{Huang}},
  \bibnamefont{et~al.}, \bibinfo{journal}{Phys. Rev. B}
  \textbf{\bibinfo{volume}{44}}, \bibinfo{pages}{3604} (\bibinfo{year}{1991}).

\bibitem[{\citenamefont{Anisimov et~al.}(1991)\citenamefont{Anisimov, Zaanen,
  and Andersen}}]{ldapu91}
\bibinfo{author}{\bibfnamefont{V.~I.} \bibnamefont{Anisimov}},
  \bibinfo{author}{\bibfnamefont{J.}~\bibnamefont{Zaanen}}, \bibnamefont{and}
  \bibinfo{author}{\bibfnamefont{O.~K.} \bibnamefont{Andersen}},
  \bibinfo{journal}{Phys. Rev. B} \textbf{\bibinfo{volume}{44}},
  \bibinfo{pages}{943} (\bibinfo{year}{1991}).

\bibitem[{\citenamefont{Czy\ifmmode~\dot{z}\else \.{z}\fi{}yk and
  Sawatzky}(1994)}]{amf_fll}
\bibinfo{author}{\bibfnamefont{M.~T.} \bibnamefont{Czy\ifmmode~\dot{z}\else
  \.{z}\fi{}yk}} \bibnamefont{and} \bibinfo{author}{\bibfnamefont{G.~A.}
  \bibnamefont{Sawatzky}}, \bibinfo{journal}{Phys. Rev. B}
  \textbf{\bibinfo{volume}{49}}, \bibinfo{pages}{14211} (\bibinfo{year}{1994}).

\bibitem[{\citenamefont{Anisimov et~al.}(1993)\citenamefont{Anisimov, Solovyev,
  Korotin, Czy\ifmmode~\dot{z}\else \.{z}\fi{}yk, and Sawatzky}}]{ldapu93}
\bibinfo{author}{\bibfnamefont{V.~I.} \bibnamefont{Anisimov}},
  \bibinfo{author}{\bibfnamefont{I.~V.} \bibnamefont{Solovyev}},
  \bibinfo{author}{\bibfnamefont{M.~A.} \bibnamefont{Korotin}},
  \bibinfo{author}{\bibfnamefont{M.~T.} \bibnamefont{Czy\ifmmode~\dot{z}\else
  \.{z}\fi{}yk}}, \bibnamefont{and} \bibinfo{author}{\bibfnamefont{G.~A.}
  \bibnamefont{Sawatzky}}, \bibinfo{journal}{Phys. Rev. B}
  \textbf{\bibinfo{volume}{48}}, \bibinfo{pages}{16929} (\bibinfo{year}{1993}).

\bibitem[{\citenamefont{Braun et~al.}(2006)\citenamefont{Braun, Min\'{a}r,
  Ebert, Katsnelson, and Lichtenstein}}]{braun}
\bibinfo{author}{\bibfnamefont{J.}~\bibnamefont{Braun}},
  \bibinfo{author}{\bibfnamefont{J.}~\bibnamefont{Min\'{a}r}},
  \bibinfo{author}{\bibfnamefont{H.}~\bibnamefont{Ebert}},
  \bibinfo{author}{\bibfnamefont{M.~I.} \bibnamefont{Katsnelson}},
  \bibnamefont{and} \bibinfo{author}{\bibfnamefont{A.~I.}
  \bibnamefont{Lichtenstein}}, \bibinfo{journal}{Physical Review Letters}
  \textbf{\bibinfo{volume}{97}}, \bibinfo{eid}{227601}
  (pages~\bibinfo{numpages}{4}) (\bibinfo{year}{2006}).

\bibitem[{\citenamefont{Hewson}(1997)}]{hewson}
\bibinfo{author}{\bibfnamefont{A.~C.} \bibnamefont{Hewson}},
  \emph{\bibinfo{title}{The {Kondo} problem to heavy fermions}}
  (\bibinfo{publisher}{Cambridge University Press}, \bibinfo{year}{1997}).

\end{thebibliography}

\end{document}